# Spectral Structures and Their Generation Mechanisms for Solar Radio Type-I Bursts

**Running Heads:** Spectral Structures of Type-I Bursts

K. Iwai[1], Y. Miyoshi[2], S. Masuda[2], F. Tsuchiya[3], A. Morioka[3], and H. Misawa[3]

1. Nobeyama Solar Radio Observatory, National Astronomical Observatory of Japan, Nobeyama, Nagano 384-1305, Japan; kazumasa.iwai@nao.ac.jp
2. Solar-Terrestrial Environment Laboratory, Nagoya University, Nagoya, Aichi 464-8601, Japan
3. Planetary Plasma and Atmospheric Research Center, Tohoku University, Sendai, Miyagi 980-8578, Japan

## Abstract

The fine spectral structures of solar radio type-I bursts were observed by the solar radio telescope AMATERAS. The spectral characteristics, such as the peak flux, duration, and bandwidth, of the individual burst elements were satisfactorily detected by the highly resolved spectral data of AMATEAS with the burst detection algorithm that is improved in this study. The peak flux of the type-I bursts followed a power-law distribution with a spectral index of 2.9–3.3, whereas their duration and bandwidth were distributed more exponentially. There were almost no correlations between the peak flux, duration, and bandwidth. That means there were no similarity shapes in the burst spectral structures. We defined the growth rate of a burst as the ratio between its peak flux and duration. There was a strong correlation between the growth rate and peak flux. These results suggest that the free energy of type-I bursts that is originally generated by non-thermal electrons is modulated in the subsequent stages of the generation of non-thermal electrons, such as plasma wave generation, radio wave emissions, and propagation. The variation of the time scale of the growth rate is significantly larger than that of the coronal environments. These results can be explained by the situation that the source region may have the inhomogeneity of an ambient plasma environment, such as the boundary of open and closed field lines, and the superposition of entire emitted bursts was observed by the spectrometer.

**Subject Headings**: Sun: corona - Sun: radio radiation

1 Introduction

The solar corona contains many particle acceleration phenomena that are caused by the interactions between the coronal magnetic field and plasma. Non-thermal electrons accelerated in the corona emit radio waves in the metric range. As a result, many types of solar radio bursts are observed (McLean & Labrum 1985, and references therein).

One of the main emission processes of metric solar radio bursts is the plasma emission. The plasma emission is emitted around a local plasma frequency. When non-thermal electrons move in a density-gradient region, a frequency drift structure appears, as observed in type-II and type-III bursts. On the other hand, there are also many types of complex fine spectral structures in the solar radio bursts that are difficult to be explained as a simple frequency drift structure (see review by McLean & Labrum 1985). They are thought to be caused by some diversities of particle acceleration, wave generation, radio emission, and radio propagation processes. The improvement of the observation technologies has discovered various additional types of spectral fine structures in the solar radio bursts.

A solar radio type-I noise storm is one of the most frequently observed solar radio phenomena at a metric frequency range. Two components are included in this phenomena; short duration (0.1–1 s) and narrowband ($\delta f \sim f \times \text{several } f$) emissions called type-I bursts and long duration (hours to days) and wideband ($\delta f \sim f$) continuum emissions. The continuum and burst emissions are emitted simultaneously. Both of them are highly circularly polarized up to 100 %. The sense of polarization is always O-mode. The observational results of type-I noise storms are summarized by Elgarøy (1977).

It is generally held that non-thermal electrons generated in the corona are trapped in a closed magnetic loop and excite the plasma waves. The plasma waves are converted to O-mode radio waves and finally observed as type-I. Although the detailed generation processes of type-I have not been well understood, Benz & Wentzel (1981) suggested that a newly emerging magnetic field interacts with a pre-existing coronal magnetic field and their reconnections could produce non-thermal electrons. Spicer et al. (1982) suggested that an emerging magnetic field may produce shocks, on which electrons may be accelerated.

It has been thought that the spectral structure of a radio burst is important for understanding its generation mechanism. In particular, peak flux distributions of radio bursts have been investigated by many works. For example, power-law peak flux distributions are observed in type-III bursts with a spectral index of 1.45 ± 0.31 (Aschwanden et al. 1998), in interplanetary type-III storms or micro type-III bursts with a spectral index of 2.1 (Eastwood et al. 2010) or 3.6 (Morioka et al. 2007), and in metric and decimetric spikes with a spectral index smaller than 2 (Nita et al. 2008; Mészárosová et al. 2000). There are other studies that show the log-normal distributions of interplanetary type-III storms (Cairns & Robinson 1998) and exponential distributions of metric and decimetric spikes (Robinson et al. 1996; Aschwanden et al. 1998; Mészárosová et al. 2000; Rozhansky et al. 2008).

Explanations of the observed peak flux distributions have also been investigated. For example, the concept of self-organized criticality (SOC; Aschwanden 2011b) is used for power-law distributions, a stochastic growth theory is used for log-normal distributions (Cairns & Robinson 1998), and an unstable velocity distribution of electron cyclotron maser emissions is used for exponential distributions (Robinson et al. 1996).

The peak flux distribution of type-I bursts has also been investigated. Power-law distributions are usually observed, and their spectral index is approximately 3 at 164 and 237 MHz using the Nançay Radio Heliograph (Mercier & Trottet 1997), 2.2–2.7 at 80 MHz using the Gauribidanur radioheliograph and polarimeter (Ramesh et al. 2013), and 4–5 at a metric range using a radio spectrometer (Iwai et al. 2013). From these results, the observed spectral index seems to be relatively larger (i.e., a softer spectrum) than that of flares such as at hard X-rays (Hannah et al. 2011), soft X-rays (Shimizu 1995), and microwaves (Akabane 1956).

Other spectral structures, such as the duration and bandwidth, have also been investigated. The classical results are summarized in Elgarøy (1977) and McLean & and Labrum (1985). Kattenberg & van der Burg (1982) studied the peak flux, duration, and bandwidth of type-I bursts and found that there are no correlations among them.

In spite of these studies, the generation processes of type-I bursts, including the origin of the soft peak flux distribution, are still unknown. This is partially because the poor resolution of the observed data. Isliker & Benz (2001) suggested that measurements with insufficient time and frequency resolutions can derive incorrect spectral characteristics. Even though increasingly more highly resolved data have been made available, the previous studies lack discussions of the association of spectral structures with their generation processes because theories behind the radio bursts have not been well developed.

AMATERAS (the Assembly of Metric-band Aperture TElescope and Real-time Analysis System; Iwai et al. 2012b) produces highly resolved spectral data in the metric range. In addition, Iwai et al. (2013) developed a two-dimensional burst detection algorithm that can distinguish each type-I burst element from complex noise storm spectra that include numerous instances of radio frequency interference (RFI). Hence, we can perform various types of spectral analysis using these data and algorithms.

The purpose of this study is to investigate the spectral structures of type-I bursts and their generation processes. The peak flux, duration, and bandwidth of the burst element are derived using highly resolved spectral data of AMATERAS through a burst detection algorithm that is improved from Iwai et al. (2013) in this study. The derived results are compared with several theoretical models to suggest the plasma processes that contribute to the burst emissions.

The instrument and data sets used in this study are described in Section 2. The data analysis methods and results are provided in Section 3 and discussed in Section 4. The paper is summarized and concluded in Section 5.

## 2 Observations and Data Sets

AMATERAS is a ground-based solar radio telescope for spectropolarimetry in the metric range. The observation band of this telescope is 150–500 MHz with a 10-ms accumulation time and a 61-kHz bandwidth. Simultaneous observations of both the left-handed circularly polarized (LCP) component and the right-handed circularly polarized (RCP) component are possible.

Three type-I events observed on January 16, 23, and 26, 2011, were used in this study. The typical burst intensity was strong on January 23, medium on January 26, and weak on January 16. Their polarizations were left-handed (January 16), right-handed (January 23), and non-polarized (January 26). Hence, this data set contains various situations of type-I bursts. Type-I bursts observed between 2:06 and 4:16 on January 16, 1:05 and 2:05 on January 23, and 3:00 and 4:00 on January 26 are used in the present data analysis. Only the prominent sides of polarizations (and RCP for January 26) are analyzed. The event observed on January 26 is analyzed briefly in Iwai et al. (2013). Figure 1 shows the dynamic spectra of these events. No flares larger than C class (Watanabe et al. 2012) or coronal mass ejections (CMEs) were reported during the observational periods on January 16 and 26[1]. Hence, the type-I modulation by flares (Aurass et al. 1990, 1993) or CMEs (Chertok et al. 2001; Iwai et al. 2012a) can be neglected. A CME was observed during the observational period on January 23. Hence, the enhancement of the typical burst intensity immediately after the beginning of the observation might be due to the CME.

## 3 Data Analysis and Results
### 3.1 Data Analysis Method

The burst search algorithm by Iwai et al. (2013) is used in this study. In this algorithm, RFIs are removed from the observed radio spectra by applying a moving median filter with a bandwidth of 610 kHz (10 bins) along the frequency axis. Burst and continuum components are distinguished by a two-dimensional maximum and minimum search of the radio dynamic spectra.

In addition to the above peak search, two fitting analyses are used to define the duration and bandwidth of the burst. A Gaussian fitting along the time axis is applied to each peak flux of the burst. The full-width-half-maximum (FWHM) of the fitted Gaussian function is defined as the duration of the burst. A Gauss fitting along the frequency axis is also applied to each peak flux. The FWHM of the fitted Gaussian function is defined as the bandwidth of the burst. In case the next burst element is inside the half maximum of the previous element, the FWHM of these two neighboring bursts cannot be defined. Hence, these bursts are eliminated from the statistics in this study. Therefore, bursts that have all of the three defined parameters (peak flux, duration, and bandwidth) are used in the following analysis. Figure 2 shows an example of the burst element that showed in Iwai et al (2013)

---

[1] http://cdaw.gsfc.nasa.gov/CME_list/

with its peak flux, and the FWHMs of its duration and bandwidth.

In Figure 2, the actual time profile of the burst is asymmetric (shorter rising phase and longer decay phase). We observed both burst elements that have symmetric and asymmetric time profiles. We also observed bursts that have longer rising phase and shorter decay phase. These results are consistent with the result in Sundaram and Subramanian (2005) that observed both symmetric (38%) and asymmetric (62% including both shorter and longer rising phases) time profiles of type-I bursts. Although there are several definitions for the duration of the burst, we fit all the bursts as a symmetric Gaussian function and define its FWHM as the duration of the burst. The merit of this definition is that we can deal with all bursts equally regardless of its shape. Because this study focuses on the statistical properties of the burst elements, our approach for the data analysis in this study should be plausible. It should be noted that the focusing on the statistical property prevents from discussing on the characteristics of the individual burst element. For the future study, more proper definition of the duration (for example, time period between the initiation and peak of the individual element) will be necessary.

### 3.2 Results

Figure 3 shows the results of the statistical study of January 26, 2011. The top left panel of Figure 3 is the peak flux distribution of the type-I bursts. The solid and dotted lines show power-law and exponential fitting curves, respectively. The peak fluxes of the bursts follow a power-law distribution. The spectral index of this power-law is greater than 3.

The middle and bottom left panels of Figure 3 show distributions of the duration and bandwidth, respectively. The solid and dotted lines show power-law and exponential fitting curves, respectively. The distributions of duration and bandwidth are more similar to exponential distributions than power-law distributions.

The top right panel of Figure 3 shows a scatter plot of the peak flux intensity vs. duration of the type-I bursts. The correlation coefficient is 0.12. Hence, there is almost no relationship. The middle and bottom right panels show a scatter plot between the peak flux intensity and bandwidth and a scatter plot between the duration and bandwidth, respectively. These figures also show small correlations.

These results suggest that the peak flux, duration, and bandwidth do not correlate with each other. In other words, there is no similarity shapes in the burst spectral structures.

Figures 4 and 5 show the same analysis results of January 16 and 23, 2011. Although the typical burst intensities are different, the results have similar trends to that of January 26.

# 4 Discussion

The peak fluxes of type-I bursts followed a power-law distribution. This distribution is consistent with previous studies (Mercier & Trottet 1997; Ramesh et al. 2013; Iwai et al. 2013). The observed spectral indexes in previous studies are about 3 (Mercier & Trottet 1997), 2.2–2.7 (Ramesh et al. 2013), and 4–5 (Iwai et al. 2013). The spectral index of our observational result is between 2.9 and 3.3. It is in the range of the previous results. There are almost no correlations between the peak flux, duration, and bandwidth.

Because we used only the bursts whose three parameters are defined, the derived spectral index on January 26 is different from that of the previous paper (Iwai et al. 2013). The bursts that are defined by these three parameters should be isolated from the other burst elements. In particular, it is difficult to provide a Gaussian fit for weak bursts. Hence, they are more often eliminated from the population of the statistical analysis. Therefore, the spectral index becomes smaller (i.e., a harder spectrum). However, this effect does not influence the following discussion because the main parts of the discussion consider the ratios and relationships between the peak flux, duration, and bandwidth of the individual burst elements and do not consider the value of the spectral index.

## 4.1 Comparison with the Avalanche Model

SOC (Bak, Tang, & Wiesenfeld, 1987) is known to produce a power-law peak flux distribution. Rosner & Vaiana (1978) proposed the avalanche model for solar flares. Subsequently, many works have proposed analytical and numerical models of power-law distributions of solar flares using this concept (e.g., Aschwanden et al. 1998; Aschwanden 2011b). We use the basic avalanche model developed by Rosner & Vaiana (1978) for simplicity. In this model, a process grows exponentially with a constant growth rate $\Gamma$ ($\Gamma = 1/\tau_G$, where $\tau_G$ is the growth time). Equation 1 show the time variation of the process (light curve of type-I burst in this study),

$$F(x) \propto \exp\left(\frac{x}{\tau_G}\right). \quad (1)$$

This process saturates after the saturation time $t$. The peak intensity of this process (= $F(t) = F_s$) is derived as follows:

$$F(t) \propto \exp\left(\frac{t}{\tau_G}\right). \quad (2)$$

The saturation times of this process are random and follow a Poisson distribution with an e-folding time $\tau_S$. The distribution of the saturation time of this process (= $N(t)ds$) is as follows:

$$N(t)ds \propto \exp\left(-\frac{t}{\tau_S}\right)ds. \quad (3)$$

Equations 2 and 3 are solved to derive the peak flux distribution of this process (= $N(F_S)$). It becomes a power-law distribution, and its spectral index $\alpha$ is derived by the ratio of the e-folding saturation time to the growth time:

$$N(F_S) \propto F_S^{-\alpha} \quad (4)$$

$$\alpha = 1 + \frac{\tau_G}{\tau_S}. \quad (5)$$

In this study, the FWHM of the burst duration is analyzed. The saturation time of the avalanche model and FWHM of the burst duration are not strictly the same. However, there is only a small asymmetry in the light curve of the type-I burst. Hence, the FWHM of the burst duration can be recognized as the similar parameter of the saturation time. In the middle left panel in Figures 3-5, the distribution of the duration time exhibits behavior that is more similar to an exponential distribution than to a power-law distribution. This distribution suggests that the duration of the type-I bursts follows a random distribution (Poisson statistics). Hence, the duration of the type-I burst can be expressed as Equation 3.

On the other hand, there should be a positive correlation between the saturation time and peak flux in the avalanche model (see Equation 2). That means the longer events should have higher intensity. However, the observational results show that there is no correlation between the observed burst peak flux and duration. Hence, the avalanche model by Rosner & Vaiana (1978) is not sufficient to explain the observed type-I burst.

There are more practical versions of the avalanche model. Aschwanden (2011b) proposed an avalanche model that has an exponential growth phase with a random saturation time and a linear decay phase that also has random decay times. Although this model does not have an analytical solution, Aschwanden (2011b) used Monte Carlo simulations and showed that this model indicates similar peak flux distributions as that of the observed hard X-ray flares in Aschwanden (2011a). In this model, the total duration is defined as the total of the saturation time and decay time. This definition is similar to that of the burst duration in this study. However, the correlation between the total duration and peak flux is also suggested in this model. Generally, the duration and peak flux should be correlated as long as the bursts have a constant growth rate (see left panel of Figure 6). Hence, there should be a similarity shapes in the burst spectral structures. Therefore, it is difficult to explain the observed type-I burst by using a model with a typical value for the growth rate.

### 4.2 Comparison with the MHD turbulence Model

The cascading of an MHD disturbance in the flare reconnection outflow (LaRosa and Moore 1993; LaRosa et al. 1994) generates electron accelerations. The spatial scale of the acceleration can form a power-law distribution. The spectral index of this power-law distribution becomes 5/3 because of Kolmogorov's law. This model is used to explain the radio spike burst, which is a metric to decimetric solar radio burst with fine spectral structures (e.g., Karlický et al. 1996; Karlický et al. 2011).

A type-I burst is thought to be emitted around the local plasma frequency. Hence, the observed frequency is converted to the electron density of the radio source region. The plasma density corresponds to the location along the density gradient, which is usually along a radial direction. Hence, the bandwidth of the burst corresponds to the spatial scale of the radio source region. The distribution of the observed bandwidth shows the exponential distribution in Figure 3-5. This result suggests that the spatial scale of the source region of a type-I burst is distributed randomly. Even though we can adopt this model to explain the peak flux distribution, the spectral index of the observed power-law peak flux distribution is between 3 and 4, which is significantly larger than 5/3. Hence, it is difficult to explain the observed type-I burst by using this model. Thus far, there seem to be no models that can explain the spectral structures of type-I bursts observed in this study.

### 4.3 Growth rate of the type-I burst

We introduce a new variable γ to define the observed spectral characteristics of the type-I bursts. γ is defined by the ratio between the peak flux and duration of the individual burst element (γ = peak flux / duration) and can be recognized as a corresponding observable of the growth rate (Γ) or growth time ($\tau_G = 1/\Gamma$) in the avalanche model. Figure 7 shows the relationship between the peak flux intensity and γ. This figure indicates that γ of the individual burst element shows large variation. In addition, there is a high correlation between the peak flux and γ of the type-I bursts. Figures 3-5 shows that a strong burst does not always have a long duration, whereas Figure 7 shows that a strong burst has a tendency to have a large growth rate γ. This result suggests that the peak intensity of the type-I burst is determined by the growth rate of the process that generates the radio burst (see right panel of Figure 6).

The slope in Figure 7 can be derived from the distribution of the duration in Figures 3-5 because the ratio between the peak flux and γ, that is the slope of Figure 7, is equal to the inverse of the duration of the burst. For example, a group of bursts that has the same duration lies on a straight line in Figure 7.

In the middle left panels of Figures 3-5, the typical durations of these three events are in the range of 100-1000s. On the other hand, the typical intensity in Figure 5 is much larger than that of Figure 4. This result also suggests that the peak flux of the burst is not determined by the duration but the growth rate.

The growth rate in this study is defined as the ratio between the peak flux and duration of the burst element. Hence, the slopes of the line between the origin and the each point in the top right panels of Figures 3-5 indicate the growth rate of the each burst element. There seems to be an upper limit of the growth rate in each type-I event. These upper limits suggest that the growth rate of each type-I event saturates at the specific level and cannot become infinite. The upper limits of the growth rates in Figures 3-5 are 470, 410, and 2800 SFU/s, respectively. Hence, the stronger type-I event seem to have larger upper limit of the growth rate.

Type-I bursts are usually associated with active regions. They are sometimes enhanced or modulated by flares and CMEs (Aurass et al. 1990, 1993; Chertok et al. 2001; Iwai et al. 2012a). The emissions of type-I bursts are usually sporadic (less than 1 s) and the emission is O-mode (e.g., Krucker et al., 1995). These facts suggest that the type-I bursts are emitted from plasma emissions that are originally generated by non-thermal electrons accelerated by some active region phenomena. In case the burst spectra are determined only by the conditions of non-thermal electrons, the spectral structures, such as the duration and peak flux, should reflect the time and spatial variations of the non-thermal electrons and their particle acceleration processes. However, the spectral index of the power-law peak flux distribution of coronal particle acceleration processes is thought to be smaller than 2, and the duration and peak flux should be correlated to at least to some degree, such as the light curves of hard X-ray flares (Aschwanden 2011b). Hence, the analysis results in this study suggest that the spectral characteristics of type-I bursts are determined by different processes than particle acceleration.

A solar radio burst phenomenon is generally thought to be composed of the generation of non-thermal electrons, plasma wave generation, mode conversion to radio emissions, and radio wave propagation processes. The growth rate of the plasma wave generation and mode conversion processes are modulated depending on the plasma environment of the source region (e.g., Li et al. 2011a, 2011b), although these processes themselves are still debated issues. Even though type-I bursts are originally generated by non-thermal electrons, the time and spatial (= spectral) variations of the radio emissions can differ from those of the seed electrons through the wave generation and radio emission and propagation processes. Therefore, there is a possibility that the final spectral structures of type-I bursts are modulated by the plasma wave generation, radio emission, and propagation processes after the seed electrons are generated in the active region.

### 4.4 Ambient plasma environment of type-I burst source region

In this section, we discuss the cause of the spectral modulation of the type-I bursts on the assumption that the spectral structures are finally determined by the growth rate $\gamma$.

Figure 8 shows the time variation of $\gamma$ in the frequency range between 173 and 179 MHz observed on January 26, 2011. The typical bandwidth of the observed type-I bursts is in the range of several MHz (see Figure 3-5). Hence, the bursts emitted between 173 and 179 MHz (6-MHz bandwidth) can be recognized as being emitted in the same frequency band. The spatial scale of the emission region between 173 and 179 MHz is estimated to be approximately 12600 km using the 10-fold Baumbach-Allen electron density model (Allen 1947). In Figure 8, the growth rate γ changes at most four times between two temporally consecutive bursts that are emitted within 1 s.

One of the candidates of the cause of the spectral modulation of the solar radio emissions is the thermal electron temperature of the radio source region, although the number of causes cannot be limited to only one. Li et al. (2011a) investigated the intensity variation of type-III bursts using a model simulation by Li et al. (2008). They set localized enhancements of the electron temperature along the trajectory of the electron beam that generate type-III bursts and found that the radio emission intensity of type-III bursts are modulated at the place that the localized enhancements of the electron temperature are set. That means the radio emission intensity can be modulated by the plasma environment even though the seed non-thermal electrons are the same.

It is unlikely that the plasma parameters, such as thermal electron temperature, change within one second. On the other hand, the plasma temperature of the active region can vary from region to region. In case the source region of a type-I burst can be spread across different plasma environments, the type-I bursts with various growth rates can be generated. Hence, one of the possible explanations is that there can be variations in the ambient plasma parameters in the source region. Then, type-I bursts with various growth rates are emitted simultaneously from the source region. Finally, all of the emitted bursts were observed by a single-dish telescope with a beam size larger than the solar disk. Type-I bursts are associated with type-III storms, that is, a group of weak type-III-like bursts observed at hectometric and kilometric range (Hanasz 1966; Bougeret et al. 1984), which suggests that the source of these two groups of radio bursts are the same, and their source region is spread across the boundary of open and closed field lines (Del Zanna et al. 2011). Such regions usually have temperature and density gradients. Therefore, our explanation does not conflict with these observational results.

The only way to discriminate the emission region of the burst using our spectral observation data is to separate the burst according to the emission frequency. The bursts emitted at different plasma frequencies are generated from different density regions that should be different regions. The frequency dependence of the growth rate is shown in Table 1. This table shows that the bursts emitted at higher frequencies (i.e., denser region) tend to have a larger growth rate. The standard deviation of each frequency band is relatively smaller than that of the total bursts except 190-196 MHz band. This result suggests a possibility that the different density regions have different typical values of the growth rate. However, their deviations are so large that it is not clear that each frequency band (i.e., each density region) has a different typical growth rate, and further investigations are required.

5 Summary and Conclusion

We investigated the spectral structures of solar radio type-I bursts using the radio telescope AMATERAS. The highly resolved spectral data enabled us to distinguish the individual elements of a complex type-I burst. The three parameters of the burst spectrum, peak flux, duration, and bandwidth, are derived by the peak search algorithm with the Gaussian fittings along the time and frequency axes for each burst element. There were both burst elements that have symmetric and asymmetric time profiles in the observed type-I bursts. However, a symmetric Gaussian fitting is applied in order to focus on the statistical properties of the burst elements. This fitting provided plausible statistical results although it prevented this study from discussing on the characteristics of the individual burst element.

The peak flux of the bursts has a power-law distribution with a spectral index between 3 and 4, whereas the duration and bandwidth are distributed more exponentially. There are almost no correlations among the peak flux, duration, and bandwidth. We introduced a variable, growth rate ($\gamma$), that is defined as the ratio between the peak flux and duration of the burst. There is a strong correlation between the growth rate and peak flux.

From these results, we explained that the type-I burst that is initially excited by non-thermal electrons can be modulated in the wave generation and radio emission stages by the ambient plasma environment around the source region. The time variation of the growth rate fluctuates significantly more than the time scale of the coronal environment. Hence, we conclude that the source region may have the gradient of the plasma environment, such as boundary of open and closed field lines, and we observed the superposition of entire emitted bursts.

This study successfully suggests the importance of plasma wave processes to understand the solar radio type-I bursts based on the observational data. The next step is to spatially resolve the source region and estimate the plasma environment of the resolved region using other ground- and space-based observations. A wideband multichannel interferometer will be an efficient tool because it can separate the emission regions that may have different growth rates. This investigation will be established by several ongoing and future missions, such as the Murchison Widefield Array (MWA; Tingay et al. 2013), the Low Frequency Array (LOFAR; van Haarlem et al. 2013), and the Square Kilometre Array (SKA; Butler et al. 2004).

## Acknowledgement
AMATERAS is a Japanese radio telescope developed and operated by Tohoku University. This work was conducted under the joint research program of the Solar-Terrestrial Environment Laboratory, Nagoya University.

**Figures and Tables**

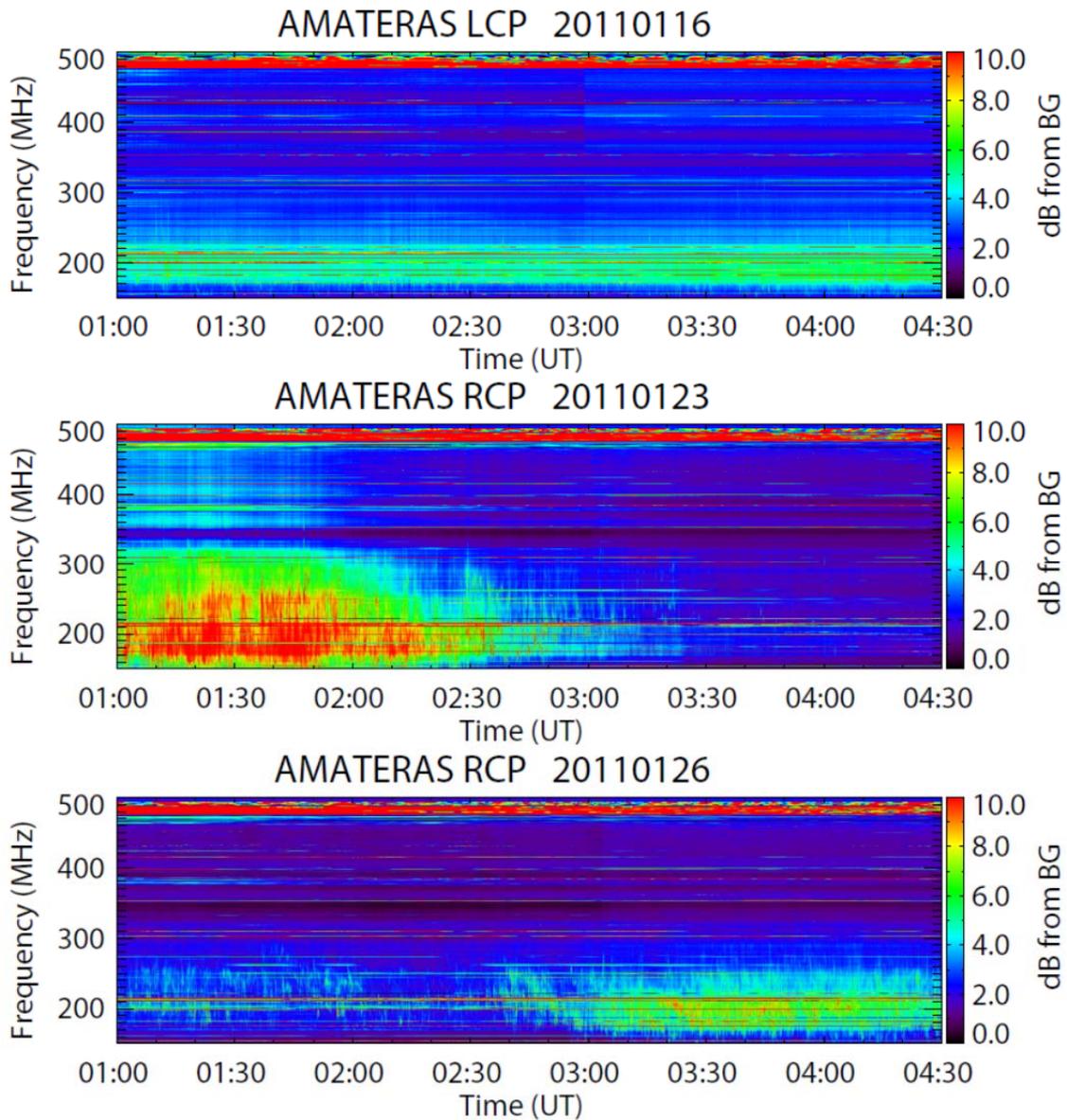

Fig. 1

Radio dynamic spectra observed with AMATERAS on January 16 (LCP, top), 23 (RCP, middle), and 26 (RCP, bottom) 2011. BG: background.

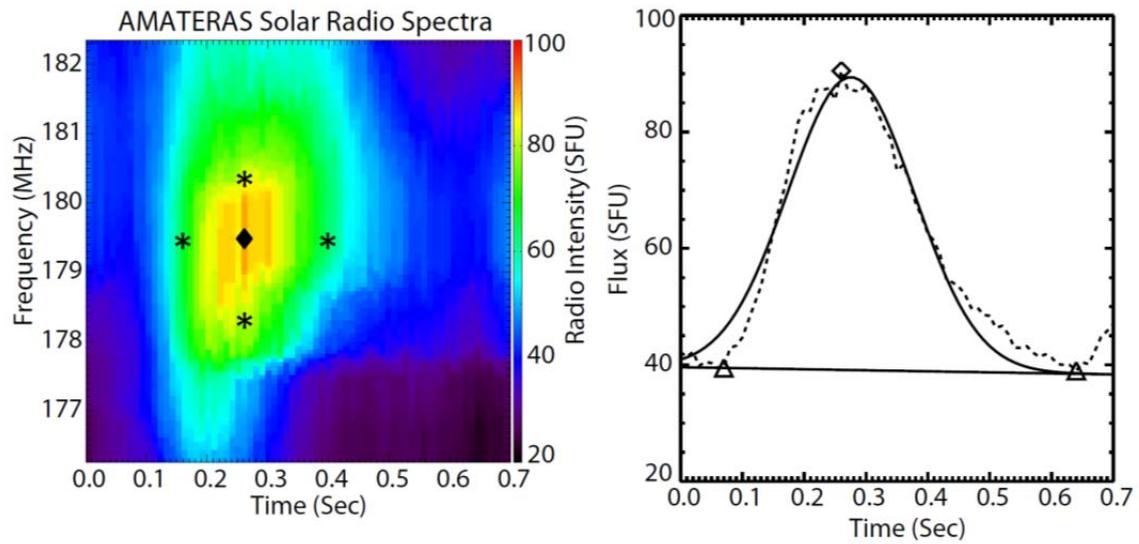

Fig. 2
(Left) An example of the radio dynamic spectra of a type-I burst element. The black diamonds denote the detected peaks of type-I bursts. The black asterisks are FWHMs of the duration and bandwidth of the burst. (Right) The observed light curve of the type-I burst includes its peak flux (dotted line) and fitted Gaussian function of the observed light curve (solid line). The diamond indicates the burst peak, and the triangles indicate the background continuum. SFU: solar flux unit[2].

---

[2] 1 SFU = $10^{-22}$ Wm$^{-2}$Hz$^{-1}$

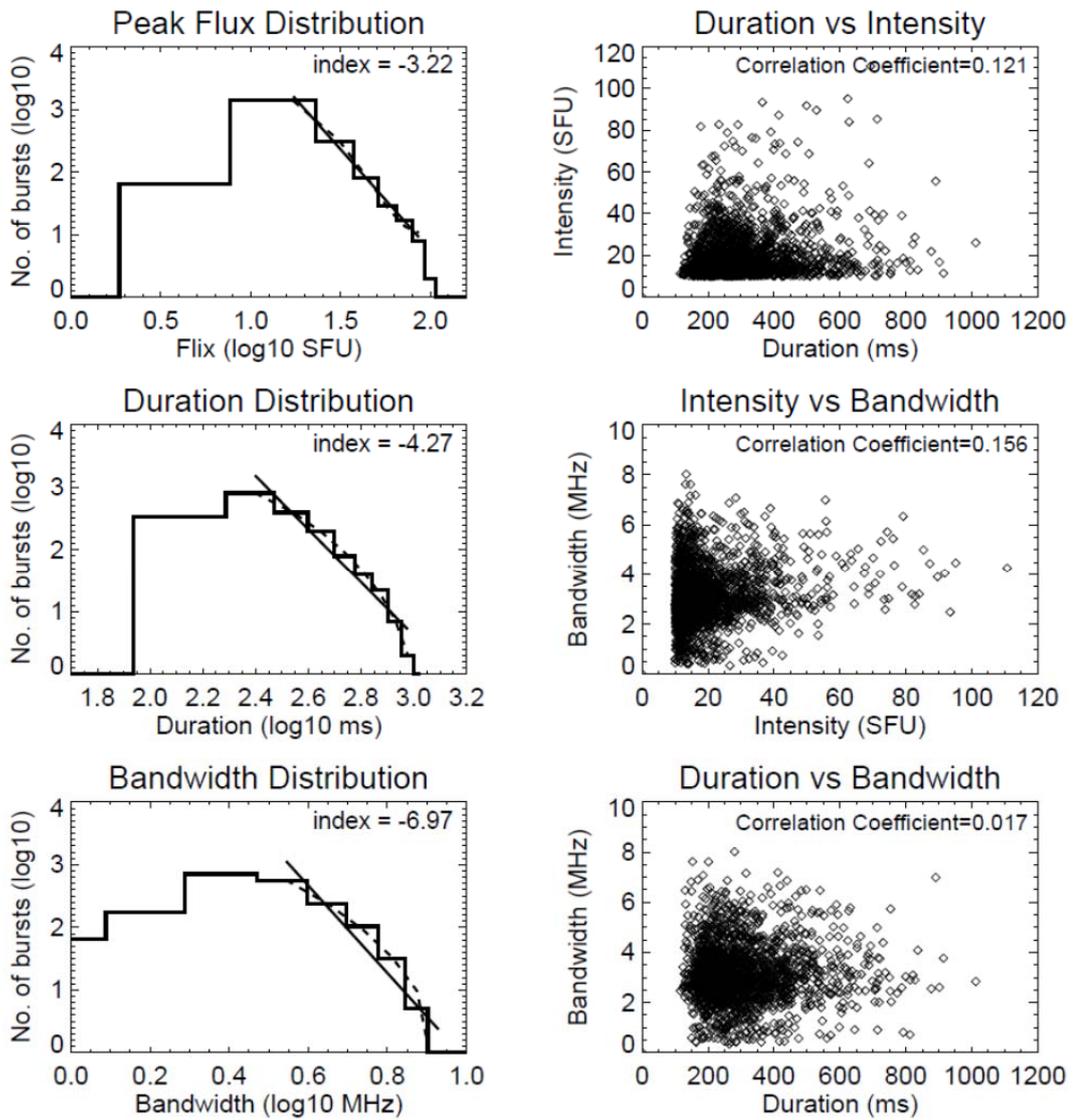

Fig. 3
Spectral characteristics of the type-I burst observed on January 26, 2011. (Top left) The peak flux distribution, (middle left) distribution of the duration, (bottom left) distribution of the bandwidth, (top right) relationships between the peak flux and duration, (middle right) peak flux and bandwidth, and (bottom right) duration and bandwidth.

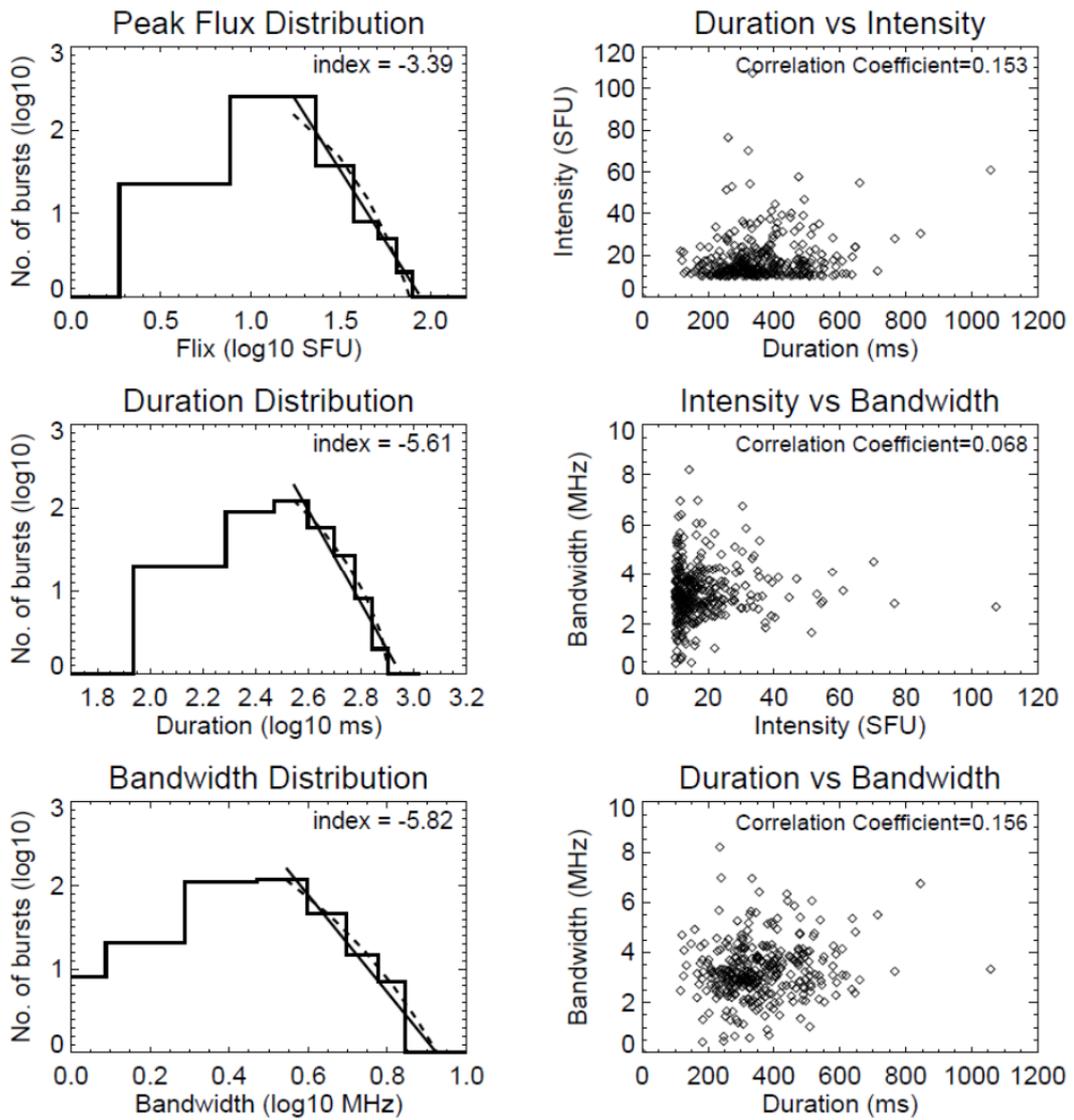

Fig. 4
Spectral characteristics of the type-I burst observed on January 16, 2011. Same format as Figure 3.

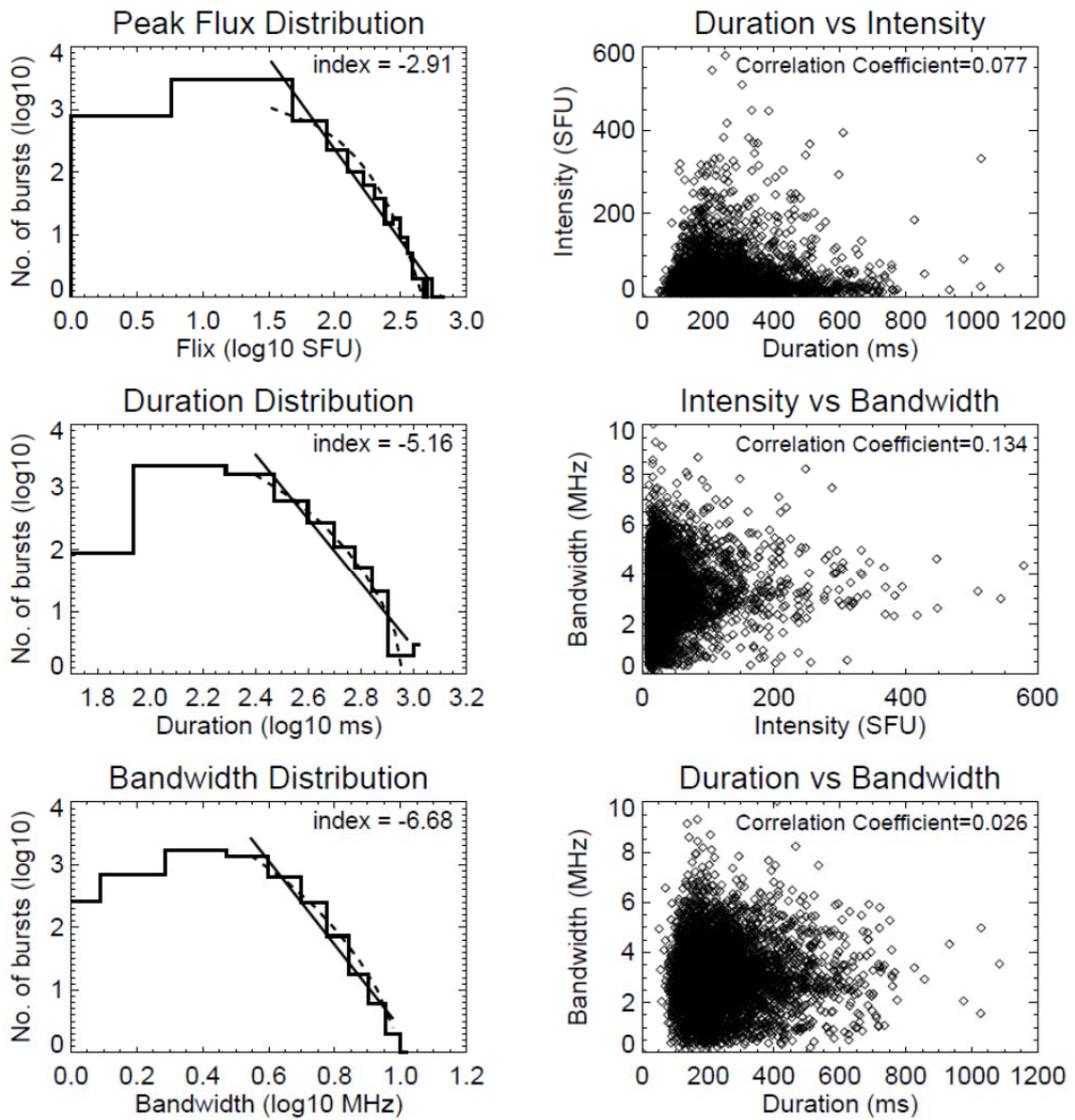

Fig. 5
Spectral characteristics of the type-I burst observed on January 23, 2011. Same format as Figure 3.

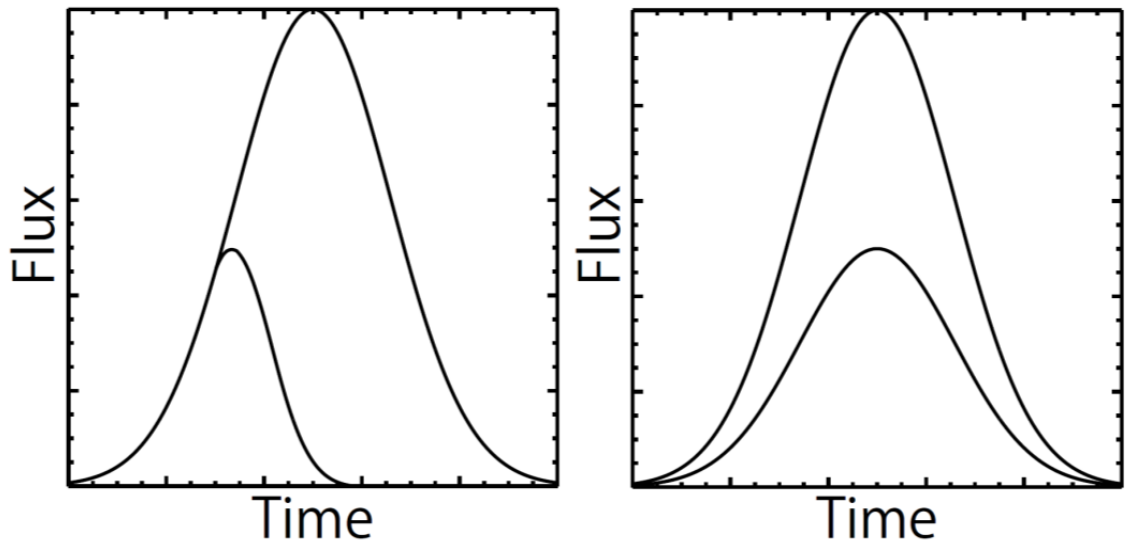

Fig. 6
(Left) Schematic image of two bursts that have the same growth rate and different saturation times. (Right) Schematic image of two bursts that have the same saturation time and different growth rates.

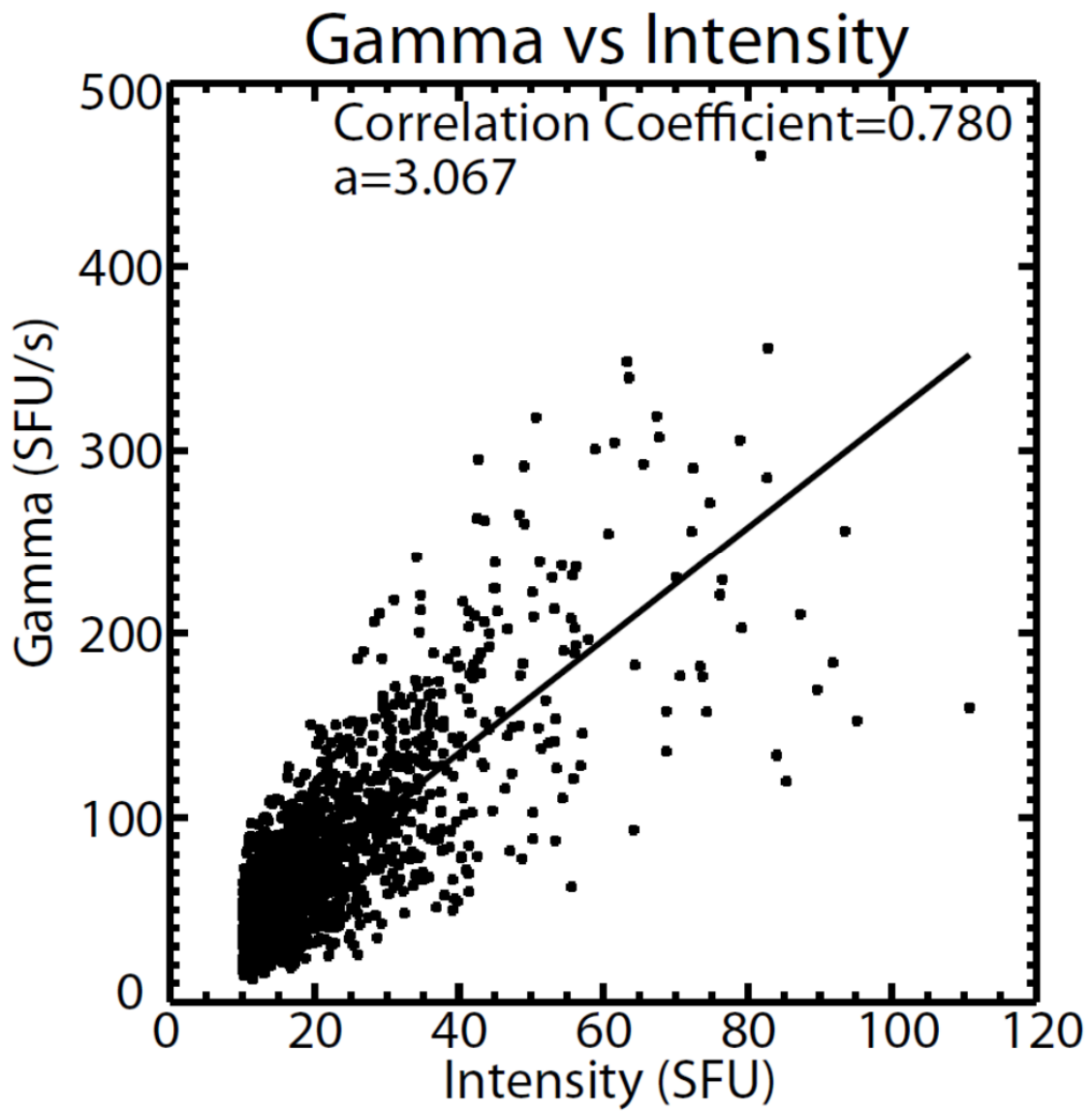

Fig. 7
Relationship between the peak flux intensity and the growth rate of type-I bursts observed on January 26, 2011.

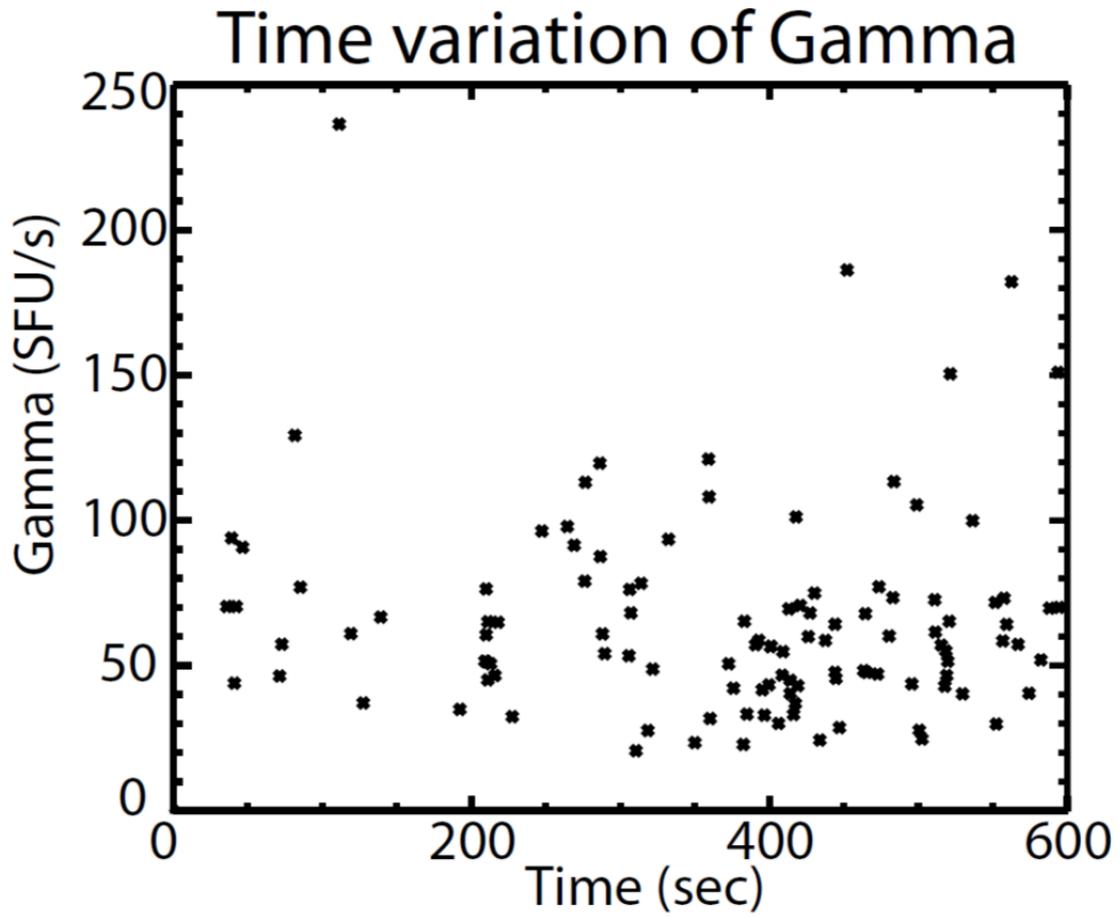

Fig .8

Time variation of the growth rates of type-I bursts in the frequency range between 173 and 179 MHz observed between 3:00 and 3:10 on January 26, 2011.

Table 1

Frequency dependence of the growth rates of type-I bursts observed between 3:00 and 3:10 on January 26, 2011.

| Frequency (MHz) | No. of elements | Average (SFU/S) | Standard deviation (SFU/S) |
|---|---|---|---|
| 172 - 178 | 113 | 65 | 35 |
| 178 - 184 | 155 | 77 | 49 |
| 184 - 190 | 169 | 72 | 42 |
| 190 - 196 | 152 | 79 | 75 |
| Total | 589 | 74 | 53 |